\documentclass[10pt,twocolumn]{article}
\usepackage[dvips]{epsfig}
\usepackage{amsmath}
\usepackage{amstext}
\usepackage{amssymb}

\def\thebibliography#1{\section*{\normalsize \bf References 
 }\list
 {[\arabic{enumi}]}{\settowidth\labelwidth{[#1]}\leftmargin\labelwidth 
 \advance\leftmargin\labelsep 
 \usecounter{enumi}} 
 \def\newblock{\hskip .11em plus .33em minus .07em} 
 \sloppy\clubpenalty4000\widowpenalty4000 
 \sfcode`\.=1000\relax}

\begin{document} 
\twocolumn[
\begin{center}
\fbox{\small PREPRINT: \hspace{1cm} Date: 1/4/1999 \hspace{1cm} Rev: 1.0 
  \hspace{1cm} Status: Physica B (\textbf{in press}) \hspace{1cm}} 
\end{center}

\begin{center} \LARGE  
   The asymmetric single-impurity Anderson model -- \\
   the modified perturbation theory
\end{center} 
 
\begin{center} \large 
   D.\ Meyer, T.\ Wegner, M. Potthoff and W.\ Nolting 
\end{center} 
 
\begin{center} \small \it  
   Institut f\"ur Physik,  
   Humboldt-Universit\"at zu Berlin, Invalidenstr. 110, 10115 Berlin,  
   Germany 
\end{center} 
\vspace{10mm} 
 
\small 

\noindent---------------------------------------------------------------------------------------------------------------------
 
\noindent{\bf Abstract} \\ 
We investigate the single-impurity Anderson model by means of the recently
introduced \textit{modified perturbation theory}. This approximation
scheme yields reasonable results away from the symmetric case. The
agreement with exactly known results for the symmetric case is
checked, and results for the non-symmetric case are presented. With
decreasing conduction band occupation, the breakdown of the screening
of the local moment is observed. In the crossover regime between Kondo 
limit and mixed-valence regime, an enhanced zero-temperature
susceptibility is found. 

\vspace{2mm}


\vspace{2mm} 
 
 
\vspace{2mm}

\noindent---------------------------------------------------------------------------------------------------------------------

\vspace{12mm} 
]
\section{Introduction}
The single-impurity Anderson model (SIAM) was originally introduced to
describe magnetic impurities in weakly correlated non-magnetic
metals~\cite{And61}. It was successfully applied to explain unexpected
low-temperature thermodynamic properties and
photoemission data~\cite{GS85,Aea86}. Although its predictions for photoemission in
periodic structures is still a matter of controversial
discussions~\cite{Aea97,Huef96,MGB96}, it remains nevertheless
surprisingly well suited for describing alloys of magnetic and
non-magnetic compounds.

The SIAM is one of the
most-examined and best-understood models of many-body physics. The
renormalization group theory~\cite{Wil75}
and
the Bethe-Ansatz (see e.g.~\cite{TW83}) have contributed
enormously to the understanding of its physics.
An excellent review of these and other theoretical approaches
to the SIAM can be found in~\cite{hewson}. 

In the recent past, the SIAM has gained much interest in the context of
the dynamical mean-field theory (DMFT)~\cite{MV89,GKKR96}. It has been shown
that in the limit of infinite dimensions, correlated lattice models can
be mapped onto single-impurity models like the SIAM. Since the Bethe
ansatz method requires special assumptions concerning the conduction
band, it is not applicable in the DMFT. Often, numerically exact methods
such as \textit{exact diagonalization} (ED) or \textit{quantum Monte
Carlo} (QMC) are used. Besides of being numerically expensive, there are
also physical limitations to these methods. ED is principally limited to
finite sized systems, while using QMC it is difficult to obtain
results for low temperatures. Both methods are not well suited for
calculating spectral densities. In addition to these methods, numerical
renormalization group calculations to both the SIAM and via DMFT to the
Hubbard model have been presented~\cite{BHP98}. This method gives by
construction very useful information for the low-energy behaviour, but
the quality concerning high-energy features remains an open question.
So there is still a need for reliable analytical methods. 

A standard approximation scheme is second-order
perturbation theory around the Hartree-Fock
solution (SOPT-HF)~\cite{BJ90}. In the context of the DMFT
it is called \textit{iterative perturbation theory}
(IPT)~\cite{GKKR96}. This method gives convincing results at special
parameter sets for the lattice problem, which map onto a particle-hole
symmetric SIAM, but does not work very well for arbitrary system
parameters. To extend the IPT to arbitrary band-fillings, an ansatz
based on the SOPT-HF self-energy was introduced~\cite{MR82,MR86,KK96}, which reproduces
the atomic limit and the Friedel sum rule~\cite{Fri56,LW60,Lan66} as well. On
the basis of this ansatz, a further improvement was recently
presented~\cite{PWN97,WPN98} using the first four moments of the
spectral density to determine the free parameters.  The method
(\textit{modified perturbation theory} MPT) works away from the
symmetric point and the magnetic properties of the
solution~\cite{PHWN98} agree well with QMC
results~\cite{Ulm98}. Since these results were obtained for the Hubbard
model within the DMFT, it is still of interest to solve the original
SIAM within the MPT. After outlining the main ideas of the MPT in the
next section, we will discuss the results for the SIAM both in the
symmetric and non-symmetric case. Comparison with exactly known results
will provide a good basis to estimate the quality of the MPT.

\section{Theory}
The Hamiltonian of the SIAM reads as
\begin{eqnarray}
  \label{eq:siam}
  \lefteqn{H=\sum_{k,\sigma}(\epsilon_k-\mu)
    c_{k\sigma}^\dag c_{k\sigma}}\nonumber
  \\ && + \sum_{k,\sigma}V_{k{\rm d}}
  (c_{{\rm d}\sigma}^\dag c_{k\sigma}
  +c_{k\sigma}^\dag c_{{\rm d}\sigma})
  \\ & &
  +\sum_\sigma (\epsilon_{\rm d}-\mu)n_{{\rm d}\sigma}
  +\frac{U}{2}\sum_\sigma n_{{\rm d}\sigma}n_{{\rm d}-\sigma}
  \nonumber
\end{eqnarray}
where $c_{k\sigma}^{(\dag)}$ annihilates (creates) a conduction band
electron with quantum numbers $k$ and spin $\sigma$, $\epsilon_k$ is the 
corresponding eigenvalue of the conduction band Hamiltonian and $\mu$ the
chemical potential. $c_{{\rm d}\sigma}^{(\dag)}$ annihilates (creates) an 
electron at the impurity-level $\epsilon_{\rm d}$.
$n_{{\rm d}\sigma}=c_{{\rm d}\sigma}^{\dag}c_{{\rm d}\sigma}$ is the
impurity occupation number operator. The hybridization strength between
conduction band and impurity electrons is given by $V_{k{\rm d}}$ and
enters all practical calculations via the hybridization function
\begin{equation}
  \label{eq:hybfunc}
  \Delta(E)=\sum_k \frac{V_{k{\rm d}}^2}{E+\mu-\epsilon_k}=V^2\int \! dx \,
  \frac{\rho_0(x)}{E+\mu-x}
\end{equation}
In this paper we assume the hybridization strength to be constant
($V_{k{\rm d}}=V$), although this is not required for the MPT. $\rho_0(E)$ is the free density of states of the
conduction electrons with bandwidth $W$.
The impurity electrons interact with each other, where the interaction
strength is $U$. Contrary to lattice models,
e.~g. the Hubbard model and the periodic Anderson model, the interaction takes
place only at one single impurity level.
\\
All relevant information can be obtained from the impurity Green function
  $G_{{\rm d}\sigma}(E)=\langle\langle c_{{\rm d}\sigma};
  c_{{\rm d}\sigma}^{\dag}\rangle\rangle_{E}$
for which we find the formal solution using the equation of motion
\begin{equation}
  \label{eq:impgfsol}
  G_{{\rm d}\sigma}(E)=\frac{1}
  {E+\mu-\epsilon_{\rm d}-\Delta(E)-\Sigma_{{\rm d}\sigma}(E)}
\end{equation}
where we introduced the impurity self-energy $\Sigma_{{\rm d}\sigma}(E)$.
The actual problem is to find an (approximate) solution for the
self-energy. We choose the modified perturbation
theory~\cite{PWN97}, 
which was applied to the Hubbard model~\cite{WPN98,PHWN98} within the
framework of the 
dynamical mean-field theory~\cite{MV89,GKKR96}.
Starting point is the following ansatz for the self-energy~\cite{KK96}:
\begin{equation}
  \label{eq:ansatz}
  \Sigma_{{\rm d}\sigma}(E)=U \langle n_{{\rm d}-\sigma}\rangle
  +\frac{a_\sigma \Sigma_{{\rm d}\sigma}^{\rm (SOC)}(E)}
  {1-b_\sigma \Sigma_{{\rm d}\sigma}^{\rm (SOC)}(E)}
\end{equation}
where $\Sigma_{{\rm d}\sigma}^{\rm (SOC)}(E)$ denotes the second-order contribution
to the SOPT-HF. The parameters $a_\sigma$ and $b_\sigma$ are chosen such that
the first four spectral moments (up to $n=3$) 
\begin{equation}
  \label{eq:moments}
  M_{{\rm d}\sigma}^{(n)}=
    \int \! dE \, E^n A_{{\rm d}\sigma}(E)
\end{equation}
are correctly reproduced.\\
$A_{{\rm d}\sigma}(E)=(-1/\pi) \Im G_{{\rm d}\sigma}(E+i0^+)$ is the spectral
density. 
This requires~\cite{PWN97}:
\begin{equation}
  \label{eq:asi}
  a_\sigma=\frac{\langle n_{{\rm d}-\sigma}\rangle(1-\langle n_{{\rm d}-\sigma}\rangle)}
  {\langle n_{{\rm d}-\sigma}\rangle^{\rm (HF)}(1-\langle n_{{\rm d}-\sigma}\rangle^{\rm (HF)})}
\end{equation}
\begin{equation}
  \label{eq:bsi}
  b_\sigma=\frac{B_{{\rm d}-\sigma}-B_{{\rm d}-\sigma}^{\rm (HF)}
    -(\mu-\widetilde{\mu}_\sigma)+U(1-2\langle n_{{\rm d}-\sigma}\rangle)}
  {U^2\langle n_{{\rm d}-\sigma}\rangle^{\rm (HF)}(1-\langle n_{{\rm d}-\sigma}\rangle^{\rm (HF)})}
\end{equation}
where the superscript (HF) denotes Hartree-Fock expectation values. 
$B_{{\rm d}\sigma}$ is a higher correlation function introduced by the
fourth moment,
which can be expressed in terms of dynamic one-particle quantities~\cite{PWN97}
\begin{eqnarray}
  \label{eq:band-shift}
  \lefteqn{\langle n_{{\rm d}\sigma}\rangle(1-\langle n_{{\rm d}\sigma}\rangle)
    (B_{{\rm d}\sigma}-\epsilon_{\rm d})=}\nonumber\\ &&
    =\sum_k V_{k {\rm d}}\langle
    c_{k\sigma}^\dag c_{{\rm d}\sigma}(2n_{{\rm d}-\sigma}-1)\rangle
  \\ & &
  =
  -\frac{1}{\pi}\Im \int \! dE \, f_-(E)\Delta(E)
  \left(\frac{2}{U}\Sigma_{{\rm d}\sigma}(E)-1\right)G_{{\rm d}\sigma}(E)
  \nonumber
\end{eqnarray}
with the Fermi function $f_-(E)=(\exp(\beta E)+1)^{-1}$.
The results from reference~\cite{KK96} are simply recovered by replacing 
$B_{{\rm d}\sigma}$ by its HF value:
\begin{eqnarray}
  \label{eq:band-shift-hf}
  \lefteqn{\langle n_{{\rm d}\sigma}\rangle^{\rm (HF)}(1-\langle n_{{\rm d}\sigma}\rangle^{\rm (HF)})
    (B_{{\rm d}\sigma}^{\rm (HF)}-\epsilon_{\rm d})=}\nonumber\\&
    =&\sum_k V_{k {\rm d}}\langle
    c_{k\sigma}^\dag c_{{\rm d}\sigma}\rangle^{\rm (HF)}(2\langle n_{{\rm
        d}-\sigma}\rangle^{\rm HF}-1)\rangle
  \\& 
  =&
  -\frac{1}{\pi}\Im \int \! dE \, f_-(E)\Delta(E)
  \left(2 \langle n_{{\rm d}-\sigma}\rangle^{\rm (HF)}-1\right)G_{{\rm
      d}\sigma}^{\rm (HF)}(E) 
  \nonumber
\end{eqnarray}
The fictitious chemical
potential $\widetilde{\mu}_\sigma$ that appears in equation
(\ref{eq:bsi}) is introduced as an additional parameter in the HF Green
function 
\begin{equation}
  \label{eq:ghf}
  G_{{\rm d}\sigma}^{\rm (HF)}(E)=\frac{1}
  {E+\widetilde{\mu}_\sigma-\epsilon_{\rm d}-\Delta(E)-U\langle n_{{\rm
        d}-\sigma}\rangle}
\end{equation}
It can be used to enforce for instance the Friedel sum rule as was done
in~\cite{KK96}, but one would restrict oneself to $T=0$. Therefore, we
choose the condition $\langle n_{{\rm d}\sigma}\rangle=\langle n_{{\rm
d}\sigma}\rangle^{\rm (HF)}$ to fix
$\widetilde{\mu}_\sigma$~\cite{PWN97}. Thus, (\ref{eq:asi}) is
always simplified to $a_{\sigma}=1$.

The modified perturbation theory constructed in that way reproduces
several non-trivial limiting cases~\cite{PWN97} such as
$U/W^{\text{(eff)}}\rightarrow 0$ ($W^{\text{(eff)}}=V^2/W$ is the
effective bandwidth of the impurity-level due to the hybridization) and
$W^{\text{(eff)}}/U \rightarrow 0$ (in the sense of~\cite{HL67}). \\ 
In the symmetric case
(i.~e. symmetric density of states, half filled conduction band and
impurity level, $\epsilon_{\rm d}=-U/2$) $b_\sigma$ vanishes, and the
ansatz~(\ref{eq:ansatz}) reduces to the SOPT-HF, since the latter reproduces the
first four moments accidently.  It is known, that the SOPT-HF in the
symmetric case gives reasonable results even for large $U$~\cite{ZH83}.

\section{Results and discussion}
\subsection{The symmetric case}
\begin{figure}[h]
  \begin{center}
    \mbox{}
    \epsfig{file=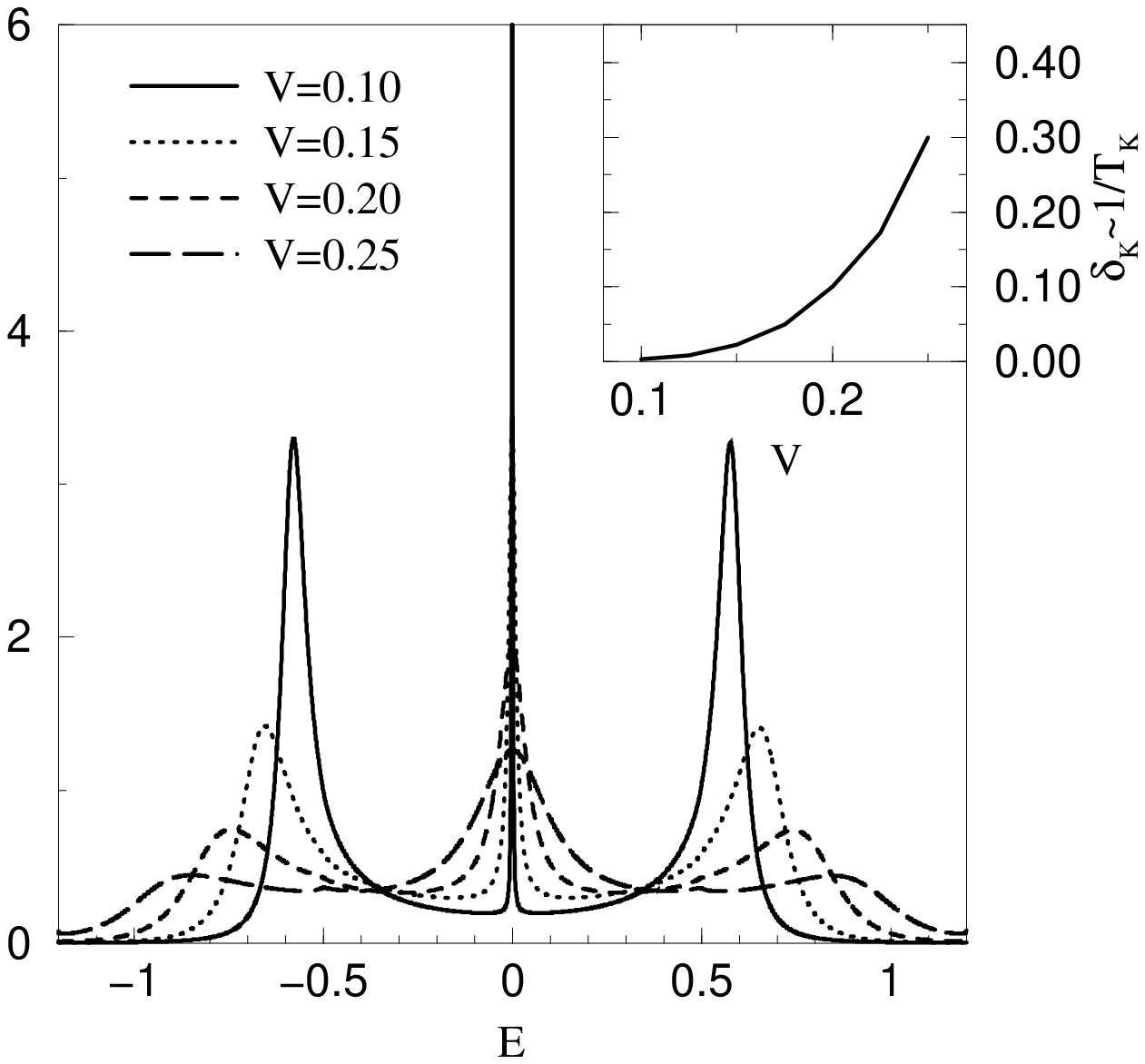, width=7cm}\\
    \epsfig{file=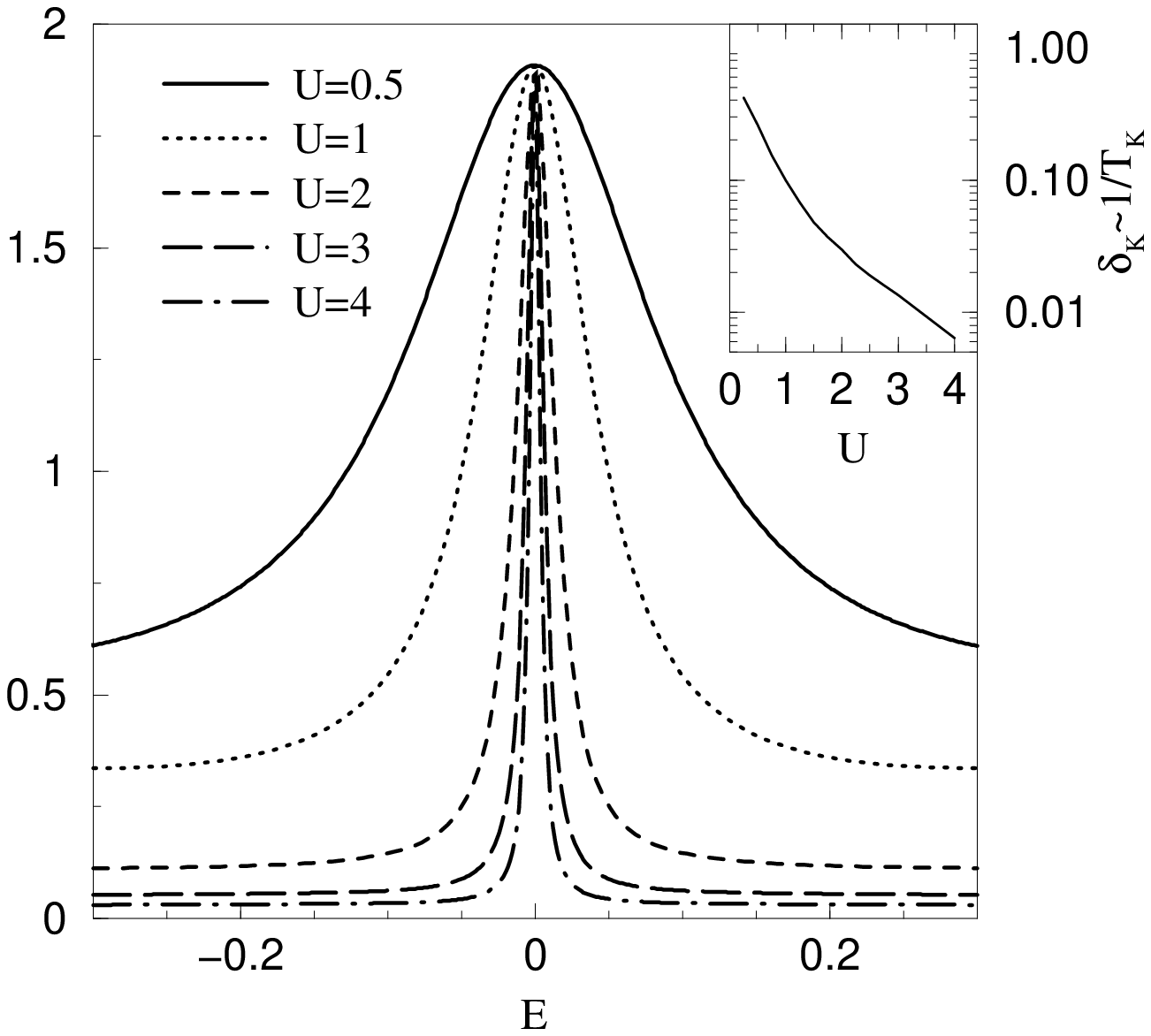, width=7cm}
    \caption{Impurity quasiparticle densities of state (impurity QDOS) of the
      symmetric SIAM ($\epsilon_{\rm d}=-\frac{U}{2}$,
      $n_{\rm d}=1.0$,$n_c=1.0$, $T=0$). Top: $U=1.0$ and different hybridization strengths $V$;
      Bottom: $V=0.2$ and various $U$. Only the vicinity of $E=\mu=0$ is 
      plotted. Inset: width $\delta_K$ of the Kondo resonance.}
    \label{fig:qdos_sym_v}
  \end{center}
\end{figure}

\begin{figure}[h]
  \begin{center}
    \epsfig{file=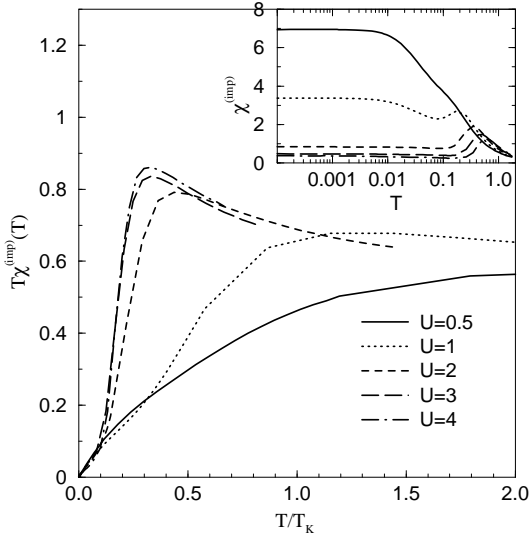,width=7cm}
    \caption{$T\chi^{(imp)}(T)$ and $\chi^{(imp)}(T)$ (inset) as
      functions of the temperature for the symmetric SIAM for different
      values of $U$ ($V=0.2$). The x-axis of the $T\chi^{(imp)}(T)$-plot
      is normalized to the respective Kondo temperature (see text).}
    \label{fig:susz_sym}
  \end{center}
\end{figure}
First we like to summarize the results for the symmetric case.
In this case, the chemical potential
will always be fixed at the center of gravity of the
conduction band. In our calculations, the conduction band density of
states is semi-elliptic and of unit width centered at $E=0$, thus
defining the energy scale used throughout this paper.  In figure
\ref{fig:qdos_sym_v} (upper picture), the impurity quasiparticle density
of states (QDOS) is plotted for various values of the hybridization
strength $V$ with constant $U=1$. For all values of $V$, it consists
of three structures. Two are located approximately at $-\frac{U}{2}$ and
$\frac{U}{2}$, respectively. These correspond to
the atomic quasiparticle levels ($\epsilon_{\rm d}$, $\epsilon_{\rm
  d}+U$) and are commonly called charge excitations.  The
central peak positioned at $\mu$ is usually denoted Kondo resonance
(KR)~\cite{hewson}. This peak is a genuine many-body effect.  The states
responsible for this feature in the QDOS are ascribed to
antiferromagnetic correlations between conduction electrons and the
localized electrons on the impurity site which form a local moment.
The antiferromagnetic correlations provide a screening of the local
moment (Kondo screening)~\cite{hewson}.  Both the width $\delta_K$ and
height of the KR scale with $V$. The inset in figure
\ref{fig:qdos_sym_v} shows the variation of $\delta_K$ with $V$.

The Kondo resonance is plotted in the lower part  of figure \ref{fig:qdos_sym_v} for
different values of $U$ and constant $V$. The height is constant, since
for the symmetric case, the MPT fulfills the Friedel sum rule as will
be discussed below. The width $\delta_K$ varies as $\sim
\frac{1}{1-{\rm const}U^2}$. 

The impurity susceptibility $\chi^{(imp)}(T)$ is defined as the
change of the susceptibility at the impurity site (denoted by $i$) due to the presence 
of the impurity:
\begin{eqnarray}
  \lefteqn{  \chi^{(imp)}(T)=}\\
  &&\frac{\partial \langle(n_{{\rm d}\uparrow}+ n_{{\rm
        c}i\uparrow}) -(n_{{\rm d}\downarrow}+ n_{{\rm
        c}i\downarrow})\rangle}{\partial B}|_{B\rightarrow0}-\chi^{(0)}_i(T)\nonumber
\end{eqnarray}
with $\chi^{(0)}_i$ being the susceptibility per lattice site of the pure system,
i.e. without the impurity, and $n_{{\rm
    c}i\sigma}=c_{i\sigma}^{\dagger}c_{i\sigma}$. 
Figure \ref{fig:susz_sym} shows $T
\chi^{(imp)}(T)$ as a function of $T$. The temperature scale is
normalized to
the Kondo temperature defined by
$T_K=\frac{1}{\chi^{(imp)}(T=0)}$~\cite{Jar95}. This quantity is proportional to
the inverse width of the KR: $T_K\sim \frac{1}{\delta_K}$~\cite{hewson}.  
For large values of $U$, a scaling
behaviour is observable, the susceptibility as function of
$\frac{T}{T_K}$ is approximately independent of $U$.
For high temperatures, $\chi^{(imp)}(T)$ becomes
Curie-like: $\chi^{(imp)}(T)\sim \frac{1}{T}$. Thus the impurity behaves like a local moment of size $T\chi^{(imp)}(T)$. For low
temperatures, strong deviations from the Curie law are obvious. For
small $U$, $\chi^{(imp)}(T)$ becomes flat, showing Pauli-like
behaviour. For $U\gtrsim 2.0$ (Kondo limit), the
susceptibility develops a maximum, the position of which scales with $T_K$ and marks the
onset of the screening of the local moment. For $T\rightarrow 0$, the
susceptibility is Pauli-like again, but with reduced value.

In summary, for the symmetric SIAM the MPT calculations, which are
equivalent to SOPT-HF in this case, agree well
with expected or exactly known results~\cite{hewson}. This is
of course due to the fact, that for the symmetric SIAM, the perturbation 
series for the self-energy
converges rapidly even for large $U$~\cite{ZH83}.

\subsection{The asymmetric case}
In this section, we present results for the non-symmetric SIAM. Away
from the symmetric point, less results are known exactly. Furthermore, 
standard perturbation theory cannot give as convincing results as in the
symmetric case. 
\begin{figure}[h]
  \begin{center}
    \epsfig{file=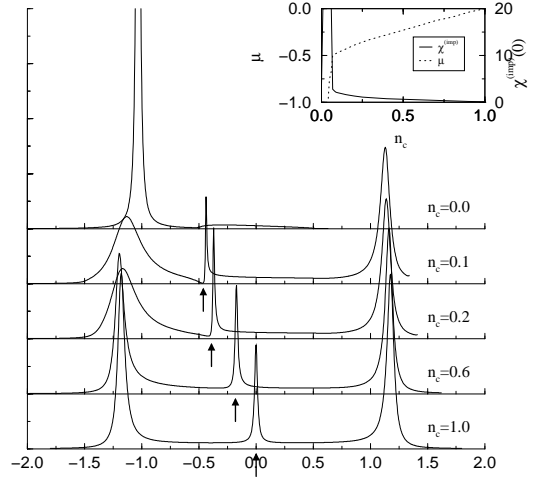,width=7cm}
    \caption{Impurity QDOS of the non-symmetric SIAM: $U=2$, $\epsilon_{\rm d}=-1.$, $V=0.2$,
      $T=0$, but various $n_c$. The chemical potential is positioned at
      the arrows. The inset shows the chemical potential $\mu$ and the
      susceptibility $\chi^{(imp)}(T=0)$ as function of the conduction
      band occupation $n_c$.}
    \label{fig:qdos_n_u2}
  \end{center}
\end{figure}

\begin{figure}[h]
  \begin{center}
    \epsfig{file=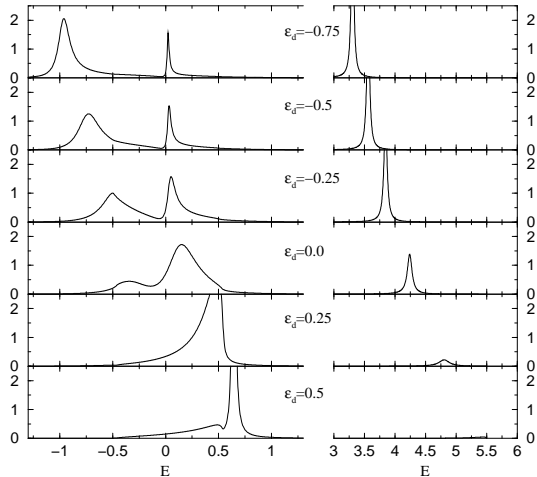,width=7cm}
    \caption{Impurity QDOS for different values of $\epsilon_{\rm d}$ and constant $U=4$,
      $V=0.2$, $n_c=1.0$ (therefore $\mu=0$).}
    \label{fig:qdos_ef}
  \end{center}
\end{figure}

First we study the variation of the conduction band occupation number
$n_c$. Starting from the symmetric case with $U=2$ and $V=0.2$, we vary
$n_c$ between $1$ and $0$. Correspondingly, the chemical potential
shifts towards lower energies. In figure \ref{fig:qdos_n_u2}, the QDOS
for different $n_c$ is plotted. The arrows indicate the position of the
chemical potential $\mu$. The inset shows $\chi^{(imp)}(T=0)$ and $\mu$
as function of $n_c$. With decreasing $n_c$, the Kondo resonance also
shifts to lower energies. Note that for $n_c<1$ the chemical potential
lies below the center of the KR, which additionally becomes rather
asymmetric. For very low carrier concentration $n_c$, the KR
disappears. For most carrier concentrations is the zero-temperature susceptibility almost constant, only a slight increase with reducing
$n_c$ is visible. At $n_c\approx 0.07$, however, $\chi^{(imp)}(0)$
diverges. This divergence is directly related to the disappearance of
the Kondo resonance. Exactly at $n_c=0.07$, the chemical potential $\mu$
crosses the lower edge of the conduction band. There are no
electrons which could form (via antiferromagnetic correlations) a Kondo
resonance, thus no screening occurs.  The still occupied impurity states
of the lower charge excitation peak form a local moment. Since now
$\chi^{(imp)}$ is the susceptibility of a stable local moment, the
divergence is to be expected. Only for $n_c=0$ the susceptibility vanishes
completely. This is, because the chemical potential shifts to $\mu
\rightarrow-\infty$, and the impurity level is therefore also empty.

Next, we like to investigate another asymmetric situation: we fix $U$ to
a large constant value of $U=4$ and vary the energetic position of the
impurity $\epsilon_{\rm d}$ with respect to the conduction band. The
hybridization strength is set to $V=0.2$ and the occupation number of the
conduction band $n_c=1$. The latter implies $\mu=0$. The situation is
visualized by means of the impurity QDOS in figure
\ref{fig:qdos_ef}. The lower charge excitation peak moves with
increasing $\epsilon_{\rm d}$ from below ($\epsilon_{\rm d}=-0.75$) into
the conduction band. The upper charge excitation at approximately
$\epsilon_{\rm d}+U$ also moves towards higher energies. Contrary to the
charge excitation peaks, the Kondo resonance remains at its position
$E=\mu=0$. For $\epsilon_{\rm d}\gtrsim 0$ the KR merges into the
lower charge excitation peak. The upper charge excitation looses spectral
weight as the impurity level becomes depopulated. For $e_{\rm d}\gtrsim 0.5$, a dip
indicates the onset of a gap within the lower charge excitation. This gap
will simply separate the charge excitation from hybridization-induced
states within the conduction band and is not a
correlation effect.\\
In particular, figure 4 shows that within the MPT the charge excitation
peaks are positioned close to the atomic quasiparticle levels as
expected. Contrary, within the SOPT-HF the positions do not
significantly depend on $\epsilon_{\rm d}$.

\begin{figure}[h]
  \begin{center}
    \epsfig{file=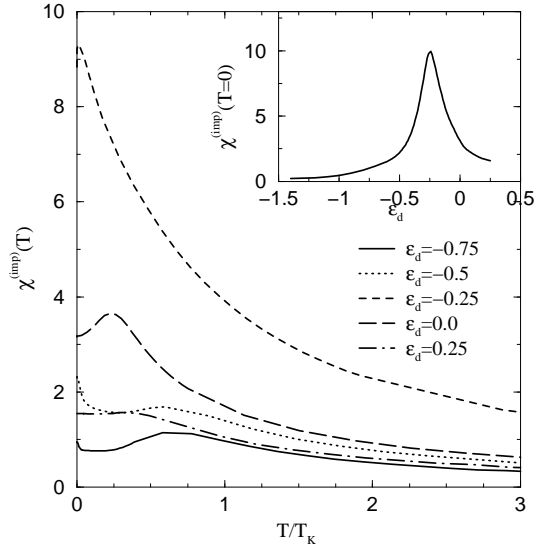,width=7cm}
    \caption{$\chi^{(imp)}(T)$ for the parameter sets of figure
      \ref{fig:qdos_ef}.}
    \label{fig:chi_ef}
  \end{center}
\end{figure}
The transition from the Kondo limit to the intermediate valence regime
(IV) and finally to the empty-impurity state manifests itself in the
susceptibility (see figure \ref{fig:chi_ef}). For high temperatures the
susceptibility follows Curie's law, $\chi^{(imp)}(T)\sim \frac{1}{T}$ in
all three cases. Only for low temperatures, $\chi^{(imp)}(T)$ behaves
differently for the respective parameter regimes.  In the Kondo limit
($\epsilon_{\rm d}\lesssim-0.5$, solid and dashed line)),
$\chi^{(imp)}(T)$ shows a maximum and decreases for decreasing
temperature.  For very low $T\lesssim 0.1 T_K$ an increase is visible,
for even lower temperatures, the susceptibility becomes constant again
and $\frac{\partial }{\partial T}\chi^{(imp)}(T)|_{T=0}=0$ (not
recognizable in figure 5).  In the IV-regime, and most pronounced around
$\epsilon_{\rm d}\approx -0.25$, the $\frac{1}{T}$-behaviour persists
down to much lower temperatures, the zero-temperature susceptibility is
enlarged, and therefore the Kondo temperature $T_K$ becomes
smaller. This can be ascribed to a weaker screening of the local
moment. A further increase of $\epsilon_{\rm d}$ depopulates the
impurity. The ``effective local moment'' shrinks and the susceptibility
gets reduced. The $T$-axis in figure \ref{fig:chi_ef} is normalized to
$T_K$ using the definition mentioned in section 3.1. No scaling
behaviour is visible.

\begin{figure}[h]
  \begin{center}
    \epsfig{file=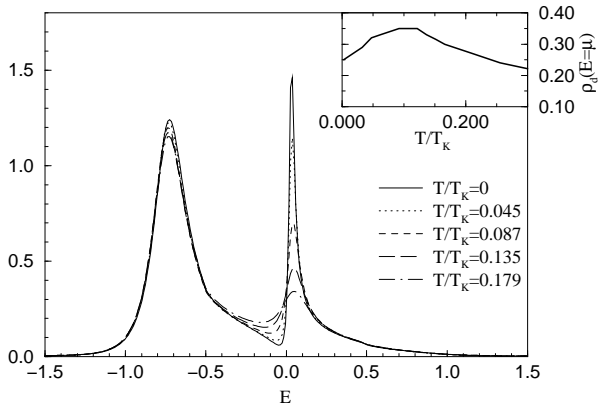, width=8cm}
    \caption{Lower part of the impurity QDOS for $\epsilon_{\rm d}=-0.5$, $U=4$, $V=0.2$ and $n_c=1.0$ at
      various temperatures. The inset shows the value of the impurity
      QDOS $\rho_{\rm d}(E)$ at  
      $E=\mu=0$.}
    \label{fig:qdos_ef_t}
  \end{center}
\end{figure}
The above-mentioned enhancement of $\chi^{(imp)}(T)$ in the Kondo regime
for low temperatures, which is most clearly observable for
$\epsilon_{\rm d}=-0.5$ needs further discussion.  This feature is
obviously most pronounced at the crossover from the Kondo limit to the
IV-regime. An interesting observation can be made by examining the
impurity density of states $\rho_{\rm d}(E)$ at $E=\mu=0$. For $e_{\rm
d}=-0.5$ (see figure 6), it shows a maximum at the same temperature
where the increase of the susceptibility begins. For $e_{\rm d}$ away
from $-0.5$, this maximum becomes less pronounced and eventually
disappears. With its disappearance the characteristic increase of the
susceptibility also vanishes. This as well as the concurrence of the position
of the maximum and the onset of the increase of the susceptibility
suggest a closer relation of the behaviour of $\rho_{\rm d}(\mu)$ and
$\chi^{(imp)}(T)$. However, it is astounding that a decrease of
$\rho_{\rm d}(\mu)$ provokes an increase in $\chi^{(imp)}(T)$.  One
possibility to explain this behaviour is the following: The states, that
build up the Kondo resonance, are responsible for the screening of the
local moment. In the examined non-symmetric case, most spectral weight
of the KR lies above $\mu$ (see figure (\ref{fig:qdos_ef_t})). Only those
electrons filling the states close to the chemical potential contribute to the
screening. If $\rho_{\rm d}(\mu)$ goes down, the number of electrons
participating in the screening decreases. The screening should become
less effective. Eventually, the local moment, which is still present,
shows a stronger reaction to an applied magnetic
field. Consequently, the susceptibility increases. In this way, the
observation may be understood. 

\begin{figure}[htbp]
  \begin{center}
    \epsfig{file=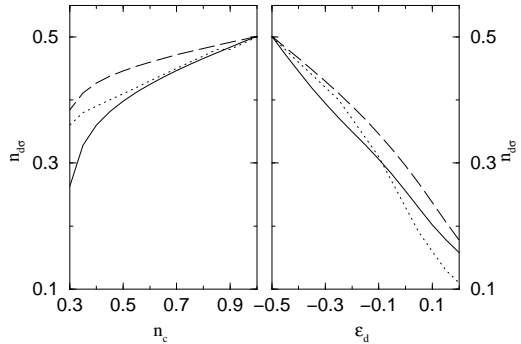, width=7cm}
    \caption{Test of the Friedel sum rule for the asymmetric SIAM. The
      impurity occupation is plotted as dotted line, the result for the
      impurity occupation using the Friedel sum rule as solid
      line (dashed line: SOPT-HF result). Left picture: $U=1.0$, $V=0.2$, $\epsilon_{\rm d}=-0.5$; right picture:
      $U=1.0$, $V=0.2$, $n_c=1.0$.}
    \label{fig:friedel_asym}
  \end{center}
\end{figure}
To finish our survey of the asymmetric SIAM, we like to check the
Friedel sum rule~\cite{hewson,Fri56} numerically. The Friedel sum rule
relates the occupation number of the impurity level, $\langle n_{{\rm
    d}\sigma} \rangle$,
to the real part of the self-energy  at the Fermi energy:
\begin{equation}
  \label{eq:friedel}
  \langle n_{{\rm d}\sigma}\rangle=\frac{1}{2}-\frac{1}{\pi} \arctan
  \left(\frac{\epsilon_{\rm d}+\Re \Sigma_{\sigma}(0)+\Re \Delta(0)}{\Im
      \Delta(0)}\right )
\end{equation}
This formula can be derived using the Luttinger theorem~\cite{LW60} as
shown in~\cite{hewson,Lan66}. 

For the symmetric SIAM, (\ref{eq:friedel}) becomes trivially fulfilled in
the MPT (equivalently in the second order perturbation theory) since
$\Re \Delta(0)=0$, $\Re \Sigma_{\sigma}(0)=\frac{U}{2}$ and
$\epsilon_{\rm d}=-\frac{U}{2}$. Therefore the impurity occupation is
always $\langle n_{{\rm d}\sigma}\rangle=0.5$. Our results for the asymmetric case
are plotted in figure \ref{fig:friedel_asym}. The dotted line is the
impurity occupation $\langle n_{{\rm d}\sigma}\rangle$ within our
approximation. The solid line
(dashed line) is obtained using (\ref{eq:friedel}) with the MPT
(SOPT-HF) self-energy. In the left picture, the conduction band occupation $n_c$
is varied from $0.3$ to $1$. The other parameters are chosen such that
the right boundary represents the symmetric case for $U=1$. As
discussed above, in the symmetric point both the MPT and the
SOPT-HF fulfill the Friedel sum rule. For $n_c\lesssim 1$, however, the
SOPT-HF deviates almost immediately, whereas in the MPT, the Friedel
sum rule is met in excellent agreement until $n_c\approx 0.55$. For
lower $n_c$, the deviation increases. The origin of this breakdown 
within our theory is not yet clarified.
The right picture shows the situation of fixed values of
$U=1$ and $n_c=1$, $\epsilon_{\rm d}$ being varied. For the left
boundary, the parameters are identical to the right boundary of the
left picture (symmetric point). Varying $e_{\rm d}$ away from the symmetric
point both the MPT and the SOPT-HF do not fulfill the Friedel sum rule. 
However, in this
case, the superiority of the MPT relative to the SOPT-HF still manifests
itself in the correct positions of the charge excitations (figure 4). The SOPT-HF 
fails to reproduce these.

\section{Conclusions}
We have analysed the SIAM within the modified perturbation theory
(MPT). For the symmetric SIAM, the MPT and therefore our results are
equivalent to the SOPT-HF. In this case, the 
SOPT-HF gives reasonable results even for large values of
$U$~\cite{ZH83}. 
For the asymmetric case, the SOPT-HF does not work very
well. This may be due to the fact that the SOPT-HF fails to reproduce the
first four moments of the spectral density. However, the MPT, which
by construction always fulfills these moments, gives reasonable results
also away from the symmetric point. Within the MPT,
the positions and weights of the charge excitations are predicted
correctly since this is
closely related to the first four moments~\cite{PHWN98}.
Furthermore, we achieve a better agreement with the Friedel sum rule
compared to the SOPT-HF.

Concluding, one can state that the MPT yields reasonable results in a wide 
region of the parameter space. Its conceptual simplicity and its
numerical feasibility qualify the MPT as a valuable complement to the
QMC method in the context of DMFT.

\begin{appendix}
\section*{\normalsize \bf Acknowledgement}
Financial support of the \textit{Deutsche Forschungsgemeinschaft}
within the project \textbf{No 158/5-1} as well as the support of the
\textit{Friedrich-Naumann Stiftung} for one of us (D.\ M.) is gratefully
acknowledged. 

\bibliographystyle{/eibe/dmeyer/lit/uuu.pott}
\bibliography{/eibe/dmeyer/lit/artikel,/eibe/dmeyer/lit/artikel.nicht.vorhanden,/eibe/dmeyer/lit/arbeiten,/eibe/dmeyer/lit/buecher}

\end{appendix}
\end{document}